\pdfoutput=1
\documentclass[english]{article}
\usepackage{geometry}
\usepackage{babel}
\usepackage{xcolor}
\usepackage{graphicx}
\usepackage{amsmath,amsfonts}
\usepackage{xcolor}
\usepackage{graphicx,indentfirst,subfigure,epsfig}
 \usepackage{varioref}
 \usepackage{wrapfig}
 \usepackage{subfigure}
 \usepackage{subfigmat}
 \usepackage{amsthm}
\usepackage{hyperref}
\hypersetup{colorlinks, citecolor=red, linkcolor=blue, urlcolor=red, breaklinks=true}
\usepackage{scrextend}
\usepackage[T1]{fontenc}
\usepackage[utf8]{inputenc}
\usepackage{tgtermes}
\usepackage{nomencl}
\usepackage{algorithm}
\usepackage{algorithmic}
\usepackage{booktabs}
\usepackage{longtable}
\usepackage{lineno}
\usepackage{epsfig} 
\usepackage{color,soul}

\usepackage{subfigmat}
\usepackage{listings}
\usepackage{color}
\usepackage{enumerate}
\usepackage{url}
\usepackage{booktabs} 

\usepackage{nomencl}
\makenomenclature

\begin{document}
\large
\title{\textbf{Preprint --- PhotoTwinVR: An Immersive System for Manipulation, Inspection and Dimension Measurements\\of the 3D Photogrammetric Models of Real-Life Structures\\in Virtual Reality}}
\author{S\l awomir Konrad Tadeja\footnote{Address all correspondence to \texttt{slawomir.tadeja@alumni.cern}}$\;^{\; \star}$, 
Wojciech Rydlewicz$^{\dagger}$, 
Yupu Lu$^{\star \ddagger}$, \\ Per Ola Kristensson$^{\star}$,  
Tomasz Bubas$^{\dagger}$, Maciej Rydlewicz$^{\dagger}$
\\ \vspace{0.3 cm} \\ $^{\star}$Department of Engineering, University of Cambridge, Cambridge, U. K., \\ $^{\dagger}$Centrum System\'{o}w Softdesk,  \L \'{o}d\'{z}, Poland, \\$^{\ddagger}$School of Aerospace Engineering, Tsinghua University, Beijing, China}
\date{}
\maketitle{}
\begin{abstract}
Photogrammetry is a science dealing with obtaining reliable information about physical objects using their imagery description. Recent advancements in the development of Virtual Reality (VR) can help to unlock the full potential offered by the digital 3D-reality models generated using the state-of-art photogrammetric technologies. These models are becoming a viable alternative for providing high-quality content for such immersive environment. Simultaneously, their analyses in VR could bring added-value to professionals working in various engineering and non-engineering settings and help in extracting useful information about physical objects. However, there is little research published to date on feasible interaction methods in the VR-based systems augmented with the 3D photogrammetric models, especially concerning gestural input interfaces. Consequently, this paper presents the PhotoTwinVR -- an immersive, gesture-controlled system for manipulation and inspection of 3D photogrammetric models of physical objects in VR. Our system allows the user to perform basic engineering operations on the model subjected to the off-line inspection process. An observational study with a group of three domain-expert participants was completed to verify its feasibility. The system was populated with a 3D photogrammetric model of an existing pipe-rack generated using a commercial software package. The participants were asked to carry out a survey measurement of the object using the measurement toolbox offered by PhotoTwinVR. The study revealed a potential of such immersive tool to be applied in practical real-words cases of off-line inspections of pipelines.
\end{abstract}

\mbox{}
\nomenclature{AR}{Augmented Reality}
\nomenclature{BIM}{Building Information Modeling}
\nomenclature{HMD}{Head-Mounted Display}
\nomenclature{HUD}{Heads-Up Daily}
\nomenclature{IA}{Immersive Analytics}
\nomenclature{VA}{Visual Analytics}
\nomenclature{VR}{Virtual Reality}

\printnomenclature
\section*{Keywords}
\noindent \textit{virtual reality; industrial visual analytics; immersive analytics; digital twin; photogrammetry; off-line inspection}


\section{Introduction}

\subsection{Research problem and suggested solution}
Recent advancements in the development and adaptation of the so-called immersive interfaces, specifically VR and AR, can help to unlock the full potential offered by 3D-reality models created using state-of-art photogrammetric technologies. Photogrammetry consists in extracting information about physical objects using images \cite{thompson1966manual}. There is a noticeable gap in the literature, described in Section 2, concerning the investigation for an appropriate interaction method in the context of VR-based systems that augment data extraction, including measurements, from digital photogrammetric models. 

This paper aims to present software and provide analyses of user-based experiments to breach this gap. The research is of relevance, as the systems that combine these two technologies are gaining more and more interest from both academia and industry. Consequently, we demonstrate PhotoTwinVR -- an immersive, gesture-controlled system for manipulation, inspection and dimension measurements of a 3D photogrammetric model of a physical object in VR. Our system allows the user to manipulate the 3D model using the user's own hands (see Fig.~\ref{fig:pinch_gesture}). The user can (i) move the model to any part of the 3D space (see Fig.~\ref{fig:models_movement}); (ii) decrease or increase the model's size (see Fig.~\ref{fig:models_resizing}); or (iii) rotate the model along a chosen axis (see Fig.~\ref{fig:models_rotation}). These manipulation methods are facilitated by the hand-tracking sensor attach on the Oculus Rift HMD as shown in Fig.~\ref{fig:oculus_gear}). Moreover, the user can take distance measurements of the model using a built-in measurement toolkit, as seen in Fig.~\ref{fig:taking_measurements}-\ref{fig:retaking_measurements}. 

In addition, we also present the two-step process for system's verification and evaluation discussed together with the accompanying results.

First, we judge our system against the well-established Nielsen's guidelines \cite{nielsen_usability_1994, nielsen_enhancing_1994}. We use Nielsen's heuristics \cite{nielsen_usability_1994, nielsen_enhancing_1994} to assess and reason about the fundamental usability and functionality of our system. 
Secondly, we run an observational study \cite{lam_empirical_2012} with a small group of three domain-expert participants. Two of them worked for \textit{Urzad Dozoru Technicznego} (UDT, \textit{eng.} Office of Technical Inspection) \cite{urzaddozorutech} which is an EU Notified Body No. 1433\footnote{UDT is \textit{a Polish state institution appointed to conduct activities aiming at the safe operation of technical equipment in Poland. The main scope of the UDT activity is technical inspection defined by the provisions of Act on technical inspection (in Poland)}~\cite{urzaddozorutech}. It also acts as the UDT-CERT Certification Body, for the certification of personnel, products and management systems. UDT has, within its structure, a separate UDT Academy Unit, responsible for the technical safety training of technical specialists, senior management and personnel operating technical equipment. The UDT is also accredited by the Polish Accreditation Centre (a member of EA) to perform grant certificates, perform inspections and testing. \cite{urzaddozorutech}.} . UDT functions based on the \textit{Act on conformity assessment and market surveillance systems}. The third expert participant had experience as a designer of industrial piping systems. 

The participants were to complete an off-line inspection of an existing pipe-rack (located in \L \'{o}d\'{z}, Poland, at the junction of Lodowa and Andrzejewskiej streets \cite{Estakada}), using the 3D photogrammetric model of this piping system, while using the measurement toolbox offered by our system. This model can be reviewed online at \cite{Estakada_Model}.

This exploratory, formative study allowed us to test PhotoTwinVR system with potential target-users for whom such a toolkit could potentially bring the most benefits. The positive reactions and comments of the study participants, together with the observation of their behaviour and analysis of qualitative data, suggest that such system potentially has a breadth of applications in an engineering domain, including the immediate applications.

\subsection{Motivation}
Photogrammetry has been defined in a number of ways, differing in the approach to the problem and concentrating on its various particular aspects \cite{Chatzifoti_Thesis_2015}, including the aforementioned \cite{thompson1966manual}, but also \cite{Photogrammetry_definition_Oxford, Photogrammetry_definition_MW, Photogrammetry_definition_Collins, luhmann2013close}. However, the most general definition has been formulated by the American Society for Photogrammetry and Remote Sensing (ASPRS).
It states that photogrammetry is an art, science and technology, which deals with collecting data on physical objects and the environment, by \textit{recording, measuring, and interpreting photographic images and patterns of radiant electromagnetic energy and other phenomena} \cite{Photogrammetry_definition_ASPRS}. History and development of photogrammetry has been described and updated in detail in several notable works and summaries published throughout the last decades, for example in \cite{thompson1966manual, 2000WolfAndDewitt, kasser2002digital, 2004MikhailAndMcGlone, schenk2005introduction, 2008BLM, luhmann2013close, BriefHsitoryOfPhotogram}. Various techniques and technologies in the field of photogrammetry have been utilized to create topographic maps, orthophoto maps, digital models of elevation and to quantify distance, measure coordinates, heights, areas and volumes \cite{ABER201023}.

Photogrammetry has been widely applied in a number of fields, For example in land surveying and topography \cite{2008BLM, 2013_Bertin, 2017_Abbaszadeh}, landslide measurements \cite{ 2016_Liu}, forestry \cite{GUYON2016249, BETTINGER201765}, rock glaciers monitoring \cite{2018Thornbush},  cultural heritage preservation and restoration \cite{miles_pitts_pagi_earl_2014, 2019_Genin}, archaeology \cite{Olga2004PhotogrammetryAA, Petti2018TheUO, soton204531}, medicine and dentistry \cite{Heike_2010, LONNEMANN2017495, CAPITAN202054} and other. 

Notable, qualitative changes in the approach to this science included the introduction of digital photogrammetry \cite{kasser2002digital, 2015-HandbookOfEmergDigiTools, Chatzifoti_Thesis_2015} and, later, algorithms for semi-automated or automated triangulation of the captured images. An interesting analysis of the state-of-art of 2006 was conducted by Seitz et al. \cite{Seitz:2006:CEM:1153170.1153518}. The driving force to enhance the performance of these algorithms was the willingness to augment the ability to produce 3D models of large areas, with the focus on 3D urban mapping, which required capturing large numbers of images \cite{labatut:hal-00834926}. However, at that time, the algorithms tended to output exceedingly complex meshes and traded high-details for loss of distinct geometric features of the 3D models. The situation led to further advancements and findings, for example, the ones proposed by Labatut et. al \cite{labatut:hal-00834926}. They introduced a novel algorithm which automatically outputs segmented model, capturing its essential geometric features. Consequently, with the commercialization of the solutions \cite{2014MCCARTHY2014175}, detailed physical models and terrain reconstruction, in the form of digital 3D models, generated using multi-image digital photogrammetry, has been made accessible to wider public.

In this paper, we deal with multi-image digital photogrammetry, with digital 3D models prepared by the means of a commercial solution, i.e. the ContextCapture \cite{ContextCapture, ContextCapture_Acute3D} software. In ContextCapture, an algorithm to automatically detect pixels, which correspond to the same physical points, is implemented. Visual data extracted from numerous images allows camera positions to be defined and a precise 3D shape of the physical object and terrain to be generated \cite{ContextCapture_Acute3D}. ContexCapture was applied in city-scale 3D mapping, as in case of Paris, Tokyo, Helsinki \cite{3Dcity-Paris}, enhancing cultural heritage preservation \cite{Acute3D_heritage}, yet also in estimating the volumes in excavations for the mining industry. Other applications included architecture and civil construction industry, for example to provide context for the design phase of an investment, to enhance design and coordination workflows enabled by Building Information Modeling (BIM) \cite{NBS_BIM}, but also to aid revitalization. An exemplary digital 3D photogrammetric model of Archcathedral Basilica of St. Stanislaus Kostka (\L \'{o}d\'{z}, Poland), created to aid such actions, is presented in \cite{Softdesk_katedra}. 

Noticeably, utilizing digital 3D photogrammetric models of reality reconstruction has gradually started being used in various industries, and for different purposes. A noteworthy possible application includes assets management in petrochemical, chemical, power and oil\&gas industries, achieved by supporting the Digital Twin of Infrastructure concept \cite{DigitalTwin, MACCHI2018790}. An example of the application in the industrial projects is presented in \cite{Softdesk_koksownia}. It is a general model of a coke plant, created with data acquired by an Unmanned Aerial Vehicle (UAV), and with the help of flight automation software \cite{DroneHarmony}. The models could be used not only for measurements, but also for multipurpose quantitative or qualitative analyses of the operation as well as condition monitoring of an asset. However, there are still obstacles to be overcome. They include legislation, safety measures on sites, presence of non-flight zones (in case of applying UAV for data acquisition), existence of hazardous areas, and the specificity of the geometry of the installations, which requires significant number of images to be taken to allow the industrial installations to be reproduced in detail. Yet, developments are made and the aforementioned issues, once overcome, may also be responsible for the success of 3D photogrammetric models, when implemented on a larger scale. They may, for example, increase the safety of the on-site workers due to the ability to perform off-line inspections and inventories.

Simultaneously, implementation of digital 3D photogrammetric models to supervise the course of various construction projects, including BIM 4D enabled workflows\cite{NBS_BIM}, is getting even more attention and is sought after by the industry. The models can be applied in the construction industry for inspecting, completing inventories before and after the investment, verifying the construction site's organization quality, controlling and predicting a timely delivery of subsequent stages of the construction project, in accordance with the schedule, to adhere with the provisions of the contract with the investor or general contractor. An exemplary model of a very small construction site in Poland is shown in \cite{Construction_model}. 

Moreover, photogrammetry starts to be widely considered as a feasible way of generating high-quality content that can significantly enhance the users experience while working with an immersive environment such as VR. For instance, Vajak et al.~\cite{vajak_combining_2017} argue that to strengthen the feeling of presence in the virtual world, an appropriate mapping between the VR and real-world is needed \cite{vajak_combining_2017}. It may be stated that VR can add another layer of realism on top of the one already given by the 3D photogrammetric models, presented with the help of a standard computer screen. Consequently, the use of photogrammetry to increase the simulation fidelity is suggested \cite{vajak_combining_2017}. 

Finally, creating easily-accessible digital 3D models and the ability to comfortably manipulate them to extract data, such as dimension, surface and volume measurements, is highly relevant. A single image or even a set of images, while carrying useful information, may influence the final interpretation of the empirical data presented due to a distinct perception of the object by the viewer \cite{Mudge:2006:NRT:2384301.2384332}.

Thus, we believe that there is merit in verifying how the proper application of VR-based systems could help to ease the usage of the photogrammetric 3D models and deliver new added-value to the end-users, across many fields and industries, with particular focus on immediate applications to off-line inspections. 

\subsection{Structure of the paper}
The structure of this paper is as follows. Section 2 offers information on the literature review concerning the usage of VR and photogrammetric 3D models. Section 3 describes the case of gestural inputs for the manipulation of the photogrammetric models in VR. Whereas, Section 4 presents a detailed description and the rationale behind the architecture of PhotoTwinVR system. Furthermore, Section 5 shows the description of the experimental verification of the system. Section 6 provides a thorough discussion of the attained results, while Section 7 formulates conclusions regarding the findings and recommendations for future work. 

\section{Related Work}
Recently, with the advances in both the photogrammetry and immersive environments, the former becomes more widely considered as a feasible way of generating high-quality content for VR. Simultaneously, as remarked in 1998 by Chapman et al.~\cite{chapman_panoramic_1998}, VR opens up a range of new opportunities for application of digital photogrammetric technologies, not only for measurements purposes. Therefore, in this review, we report both on research that emphasizes enriching user-experience and facilitating the use of the VR immersive environment by the application of photogrammetric models, and on the usage of VR to aid extracting relevant data from the digital photogrammetric models, for practical, domain-related purposes.

\subsection{Photogrammetry and VR}
Research in which both the VR and photogrammetry were coupled together includes various areas of engineering. For instance, \textit{MOSIS (Multi-Outcrop Sharing \& Interpretation System)} developed by Gonzaga et al.~\cite{gonzaga_immersive_2018} aims to leverage the benefits given by an immersive environment in using photogrammetry for geoscientists working in the oil\&gas industry. MOSIS\cite{gonzaga_immersive_2018}, similarly to our system, leverages VR environment to facilitate the usage of photogrammetric models, by allowing the user to take distance measurement. It also offers several additional tools such as a built-in compass or capability to calculate the distance between the points or orthogonal distance between calculated planes \cite{gonzaga_immersive_2018}. The authors reflect on the fact that it is hard to capture the levels of immersiveness felt by the users, hence they measured the usability using \textit{System Usability Scale} \cite{gonzaga_immersive_2018}. Liu et al.\cite{liu_construction_2009} discuss the benefits of integrating UGIS with digital photogrammetry and remote sensing with interfaces such as VR to enhance urban planning and management \cite{liu_construction_2009}. Murano et al.~\cite{murnane_virtual_2019} proposed using market-available VR hardware to simulate robotic sensor data such as audio, video and motion sensors data, to facilitate research in the area of human-robot interaction. For this type of study, the simulation must reassemble as closely as possible an existing real-world environment \cite{murnane_virtual_2019}. Therefore, photogrammetry was applied to acquire data used to build their system.

Another domain attempting to leverage the photogrammetry and VR is heritage preservation and archaeology. Photogrammetry is applied specifically to reconstruct, in detail, single objects or areas of interest. For instance, Calvert et al.~\cite{calvert_design_2019} presents the \textit{Kokoda VR} system that uses the photogrammetry to provide a more immersive learning environment for high-school students. The researchers created a \textit{Virtual Reality Learning Environment (VRLE)} using a mixture of photographs, animations and photogrammetry scans \cite{calvert_design_2019}. The study concluded that there was a significant increase in students engagement while using VR compared to a group using 360$^{\circ}$ videos \cite{calvert_design_2019}. Borba et al.~\cite{borba_itapeva_2017} describes \textit{Itapeva 3D}, a system that tries to provide its user with fully immersive, VR-based environment for cyber-archaeology \cite{borba_itapeva_2017} developed using the Unity game engine \cite{unity}. Moreover, Ipteva 3D uses a \textit{target-based travel metaphor}~\cite{borba_itapeva_2017} i.e. the change in the user's viewpoint is triggered by focusing the user's view on a target point within the environment \cite{borba_itapeva_2017}. The authors also report on positive results concerning the feel of immersion observed among participants of a small user study conducted with their system \cite{borba_itapeva_2017}. The Itapeva Rocky Shelter archaeological site is also explored using the \textit{ArcheoVR} software package presented in Borba et al.~\cite{borba_archeovr:_2017}. Woods et al.\cite{woods_beacon_2018} used photogrammetry to generate detailed 3D models of graves used in a virtual simulation of the Beacon Island by the \textit{Beacon Virtua} system \cite{woods_beacon_2018}. Antlej et al.~\cite{antlej_real-world_2018} investigated the use of VR and photogrammetry-generated models in an immersive system that allows the users to ``visit'' a paleontological digging site in Victoria, Australia \cite{antlej_real-world_2018}. Juckette et al.~\cite{juckette_using_2018} looked into combining detailed, photogrammetric models with virtual surroundings constructed with the help of GIS and other data \cite{juckette_using_2018}. The authors reflected on a fact that such setup can enhance the user's experience \cite{juckette_using_2018}. Huang \cite{huang_pieces_2019} presents a VR-based serious game that uses artefacts reconstructed using the photogrammetry combined with a hand reconstructed Maya city of Cahal Pech. Sanders~\cite{sanders_neural_2018} discusses the \textit{CUNAT (the  CUNeiformAutomated Translator)} software package for automatic translations and contextualization of ancient Mesopotamian texts in a phone-based VR environment that uses photogrammetry scans of these documents for the content. Whereas Giloth et al.~\cite{giloth_vr_2017}, using photogrammetry reconstructed and visualized some object for their VR app of a no longer existing garden labyrinth of Versailles. Ad\~ao et al.~\cite{adao_bringing_2017} discuss a system that combines photographic and geographic data obtained through unmanned aerial systems (UAS) imagery with procedural modelling to generate detailed models later on immersed in a VR environment. Fritsh et al.~\cite{fritsch_gyrolog_2018} introduces the \textit{Gyrolog  Project} whose goal is to digitize a large collection of gyroscopes in a form of 3D models accessible through a VR interface using, among others, the photogrammetry method\cite{fritsch_gyrolog_2018}. See et al.\cite{see_tomb_2018} describes the room-scale VR experience in which the photogrammetry was used to reproduce the tomb of Sultan Hussein Shahin in Malacca. Another tomb excavation site, the Dayr al-Barsha in Egypt, was digitized and placed in a web-based VR environment by Lima et al.~\cite{lima_tls_2018}. Fritsch et al.\cite{fritsch_3d_2017} discuss the potential of using 3D photogrammetry, among other spatial data acquisition methods, to generate high-quality urban content used in 3D and 4D apps for VR \cite{fritsch_3d_2017}. Moreover, in Xiao et al.~\cite{xiao_geoinformatics_2018} the authors discuss geoinformatics and spatial technologies such as photogrammetry or VR as a means of protection and preservation of cultural heritage.

\section{Gestural Input as Means of Photogrammetric Models Manipulation in VR}
Noticeably, there are several publications describing research about the advantages and benefits of using photogrammetry to enrich the experiences of VR users and to apply VR to ease the usage of photogrammetric models. However, to the best of the authors' knowledge, few papers are dealing specifically with the interactive aspect of the photogrammetry-augmented, VR-based environment. It may be argued that such conjunction is one of the key components of immersive visualization. 

Investigations into gestural input for VR has a long history, for example, Nishino et al.~\cite{nishino_interactive_1997}, in his paper from 1997, introduced the \textit{TGSH (Two-handed Gesture environment SHell)} interface. Furthermore, Song et al.~\cite{song_developing_2000} discusses the \textit{Finger-Gesture} method for selection and manipulation, whereas Latoschik \cite{latoschik_gesture_2001} presented a VR designated framework for gesture detection and analysis. 

In our literature review, only a modest number of publications investigating the gestural input as the main interaction method for systems combining both VR and photogrammetry was found. For instance, See et al.~\cite{see_tomb_2018} ran a usability study with 18 participants using virtual hands facilitated by hand-held controllers to increase the feeling of presence in the VR environment. In their experiment, the authors compared two conditions in which the participants were seeing either virtual hands or controllers avatars, as they are seen in the real-world \cite{see_tomb_2018}. The study concluded that overall, using even low fidelity virtual hands is more favourable \cite{see_tomb_2018}. To remedy this issue, Zhang et al.~\cite{zhang_excontroller:_2018} proposed to augment the VR controllers with an attached, near-infrared camera to track and display the hand and finger postures. Juckette et al.~\cite{juckette_using_2018} mention the use of Leap Motion device \cite{leapmotion} to facilitate the gesture-interface for the VR experience exploring an ancient Maya temple \cite{juckette_using_2018}. Manders et al.~\cite{manders_interacting_2008} proposed their technique that leverages the fact that HMDs leave the user's chin visible for tracking. Coupled with stereo camera data to capture the user's gestural input, it can be used to manipulate 3D objects in virtual space. For comparison, our system adds only one sensor i.e. Leap Motion \cite{leapmotion}, attached to the front of the HMD and provides an alternative set of single and double-hands gestures supported by the Leap Motion SDK \cite{leapmotion}. Furthermore, by using the Leap Motion \cite{leapmotion} controller, one can build a system for recognition of more complex gestures, for instance, Clark et al.~\cite{clark_system_2016} developed real-time hand-gesture recognition system for VR using machine learning methods.

Another interesting area of research is how the appearance and fidelity of hand avatars in VR affect the user. In addition to the previously discussed studies \cite{see_tomb_2018, zhang_excontroller:_2018}, Lin et al.~\cite{lin_need_2016} investigated how different virtual hand model impact the user's body ownership illusion (BOI). He reflected, that, although the results were inconclusive, using more realistic human-hands model might be favorable \cite{lin_need_2016}. 

Moreover, Tecchia et al.~\cite{tecchia_im_2014} describes a system that utilizes  color-coded thimbles (for thumb and index fingers) tracked by the RGBD camera, i.e. RGB camera with additional depth sensor, attached to the headset for real-time projection of the user's hands and body in the Mixed Reality (MR) environment and hypothesise that such realistic hand-models can strongly impact the user's feel of presence \cite{tecchia_im_2014}.

\section{Supporting Interaction with Photogrammetric Model in VR}
\begin{figure}[!t]
    \centering
\includegraphics[width=\textwidth]{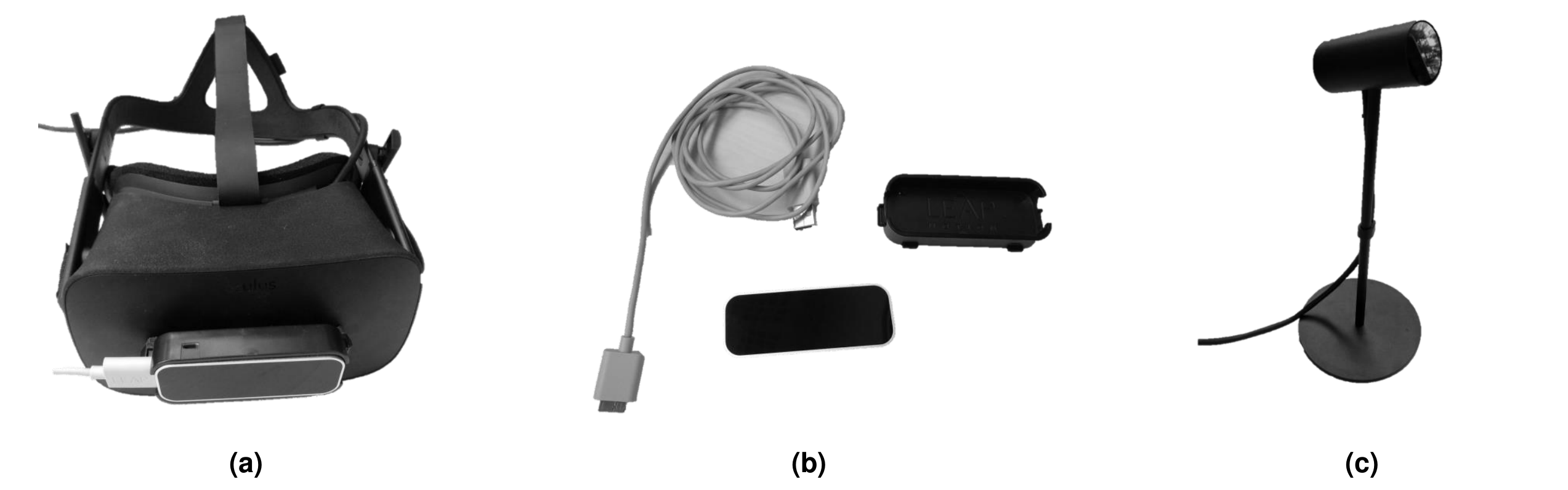}
    \caption{The visualization and tracking hardware used in our study: (a) the Oculus Rift with mounted Leap Motion; (b) Leap Motion hand-tracking sensor with mounting case and a USB cable; (c) the Oculus Rift's tracking sensor. All three components have to be connected to a supporting PC via USB (a-c) and HDMI (a) cables.}
    \label{fig:oculus_gear}
\end{figure}
The latest wave of the head-mounted displays (HMDs) can be considered as a relatively new technology concerning the previous generations of immersive interfaces. The concept of an HMD was first introduced in 1968 by Ivan Sutherland~\cite{sutherland_head-mounted_1968}. Different approaches to how to immerse the user into the VR environment, the CAVE~\cite{Cruz-Neira:1992:CAV:129888.129892}, was proposed in 1992 by Cruz-Neira et al. Presently, a plethora of HMDs exist on the market, including families of devices such as HTC Vive \cite{vive} or Facebook's Oculus \cite{oculus} (see Fig.~\ref{fig:oculus_gear}). As such, the appropriate interaction methods within VR that can be utilized across all the possible applications have to be yet fully understood. For instance, the user cannot comfortably use the standard input/output devices such as the computer mouse or keyboard when wearing the headset that -- by design -- tries to completely cut off the user's peripheral vision to intensify the feeling of immersion. 

\begin{figure}[!ht]
    \centering
\includegraphics[width=\textwidth]{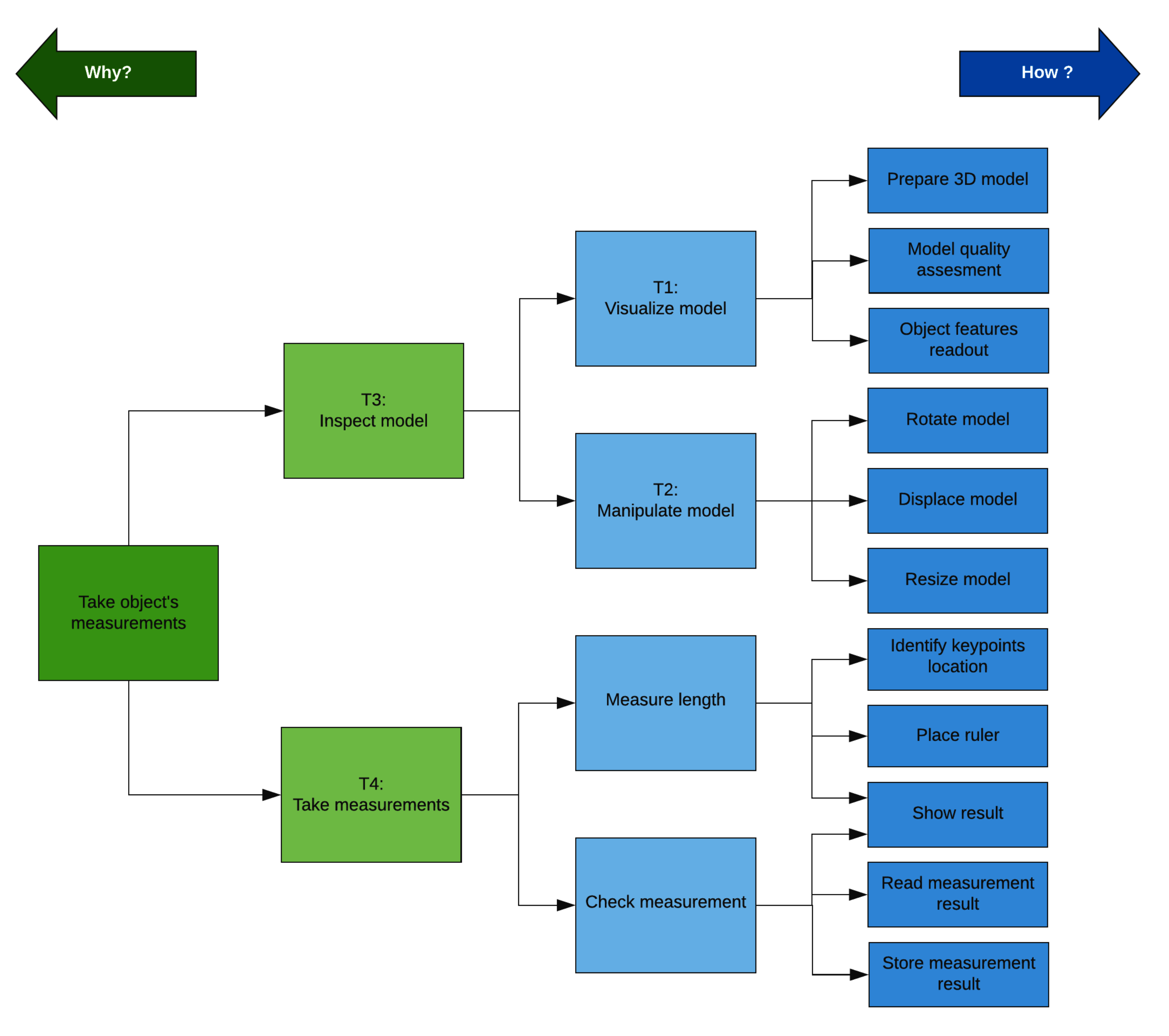}
    \caption{The FAST diagram of the function structure for the PhotoTwinVR system}
    \label{fig:fast_diagram}
\end{figure}

\subsection{Function structures} \label{FAST}
To capture the list of requirements that our system, the \textbf{PhotoTwinVR} tool, must fulfill to be effective, we employed the  \textbf{Function Analysis System Technique} (FAST) \cite{Shefelbine_edc}. The results of this analysis are presented in the form of a diagram shown in Fig.~\ref{fig:fast_diagram}. The diagram should be read from left to right in the top to bottom fashion. We start with the most high-level functionality (functions) and whilst moving in the horizontal direction, by answering the \textbf{how?} question, we split these functions into logical steps (sub-functions to the right) that are necessary in order to achieve the goal of the previous-level function (to the left) that would answer the \textbf{why?} question if we backtrack our steps. As we are looking into capture the solution-independent system requirements, we finish this process before we start to come up with actual solutions.

\subsection{Tasks Analysis}
The FAST analysis presented in the previous section \ref{FAST} allowed us to identify a list of four main, high-level tasks that user has to be able to complete using our system. Here, we discuss and describe in detail each of these primary tasks:

\textbf{T1---Model visualization}: The system should allow the user to visualize any given 3D model (e.g. a photogrammetric or CAD model) representing a 3D structure (see Fig.~\ref{fig:pipe_estacada_model}) or another 3D object which user wants to work within the VR environment.

\textbf{T2---Model manipulation}: The user should be able to effortlessly and effectively manipulate the visualized model. The manipulation functionality should allow these three basic operations: (a) rotation of the model around a chosen axis (see Fig.~\ref{fig:models_rotation}); (b) displacement of the model into a chosen position within the 3D space; (c) changing the size of the model (see Fig.~\ref{fig:models_resizing}). This functionality is facilitated with the hand-gesture tracking and recognition system i.e. the Leap Motion \cite{leapmotion} and Leap Motion SDK \cite{leapmotion}.

\textbf{T3---Model inspection}: The system should allow the user to easily visually inspect all the models' elements. This can be achieved twofold through an adequately prepared visualization (see Fig.~\ref{fig:models_rotation}) and appropriate interaction methods built-in into the system (see Fig.~\ref{fig:pinch_gesture}). Moreover, this task is influenced and vastly overlaps with task \textbf{T2} as the interaction, especially in an immersive environment such as VR, is an inseparable part of the visualization itself. Thus, the model manipulation methods take a crucial role in how the user can inspect a 3D model visualized in our system. 

\textbf{T4---Model measurements}: The system should permit the user to take and, if needed, retake desired measurements of the entire model and all its elements (see Fig.~\ref{fig:markers_creation}-\ref{fig:retaking_measurements}). Moreover, such measurements should be automatically stored and easily retrievable for future reference (see Fig.~\ref{fig:taking_measurements}).

\subsection{System Structure}
\begin{figure}[!t]
    \centering
\includegraphics[width=\textwidth]{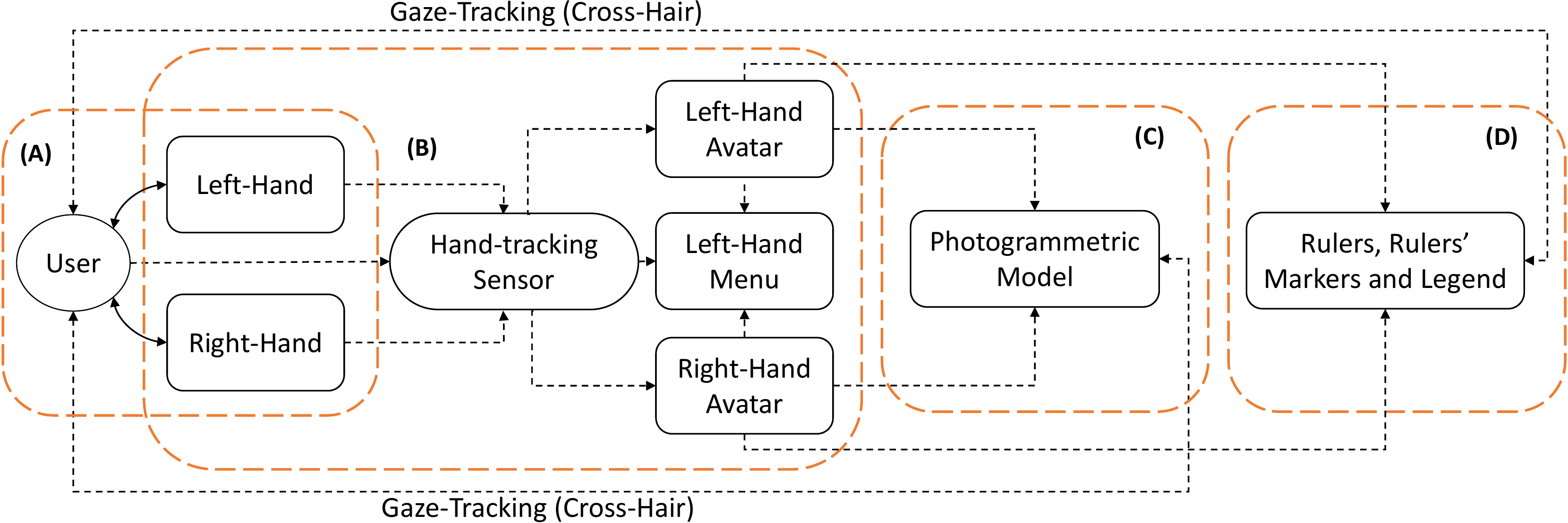}
    \caption{The diagram models the signals flowing between these four main elements of the visualization: (A) the user; (B) the hand-tracking setup; (C) the 3D model and (D) the measurement toolkit. Both the unidirectional and bidirectional signals alike are shown. Moreover, components of (A) and (B) overlap as the user's hands are both parts of the hand-tracking (B) and the user (A).}
    \label{fig:system_structure}
\end{figure}
To reason about the system structure, we decided to observe the way signals flow within the system. A similar approach was utilized in Tadeja et al.~\cite{tadeja_rae_2020} to describe the signals flux observed in the VR-based system for aeronautical design. We observed the four key components: (A) the user; (B) the hand-tracking sensor together with the avatars of the hands; (C) the photogrammetric model of an object or structure; (D) the measurement's tools. Some parts of the visualization were encompassed together as they form logical structures, and others, especially the user's hands, can be assigned to multiple groups (see Fig.~\ref{fig:system_structure}). 

\begin{enumerate}[(A)] 
\item \textbf{The user:} When speaking about signals flowing in any interactive system, the user has to be considered as one of the crucial elements as the system itself relies on receiving and sending signals to and from the user. In our case, it is especially important as the interaction is facilitated by a mixture of a gaze-tracking coupled with the hand-tracking and gesture recognition capabilities. As such, the user was a part of components (A) and (B) detailed below. The user was interacting with the system twofold. First, with the bidirectional signal via the ray-traced by an orange cross-hair, which signaled to the system what the user is focusing on at the moment. Whilst the user was gazing over an interactive element, mainly the ruler's markers (see Fig.~\ref{fig:taking_measurements}), the system responded by highlighting the object i.e. signaling to the user that this element can be interacted with. Then, the user could use a hand-gesture (see Fig.~\ref{fig:pinch_gesture}) by invoking an action dependent on both the gesture type and the object itself. Moreover, any kind of manipulation of a model, as can be seen in Fig.~\ref{fig:models_movement}-\ref{fig:models_resizing} and, i.e. its displacement, rotation and re-sizing, are signaled to the user by immediate system response. The same can be said about the operations of taking and retaking measurements of the model as shown in Fig.~\ref{fig:snapping_grid}-\ref{fig:retaking_measurements}. Hence, all the signals flowing between the user and the system are bidirectional.

\item \textbf{Hand-tracking:} As the main method of interaction is hand-tracking facilitated by a Leap Motion\cite{leapmotion} sensor, the signal received and sent by this component are usually the result of a constant feedback loop. The hands' current position and gesture are always signaled to the user via virtual avatars of hands (see Fig.~\ref{fig:pinch_gesture}-\ref{fig:markers_creation} and Fig.~\ref{fig:placing_markers}-\ref{fig:retaking_measurements}). If the hands are not visible to the user, it means that they are outside of the sensor field of tracking and as such cannot interact with the system. However, actions executed by the user on either the model  (C) or the measurement tools (D) are immediately applied and as such signaled to the user via a range of visual clues. For instance, the ruler's markers are ready to be linked with the help of ruler or moved with the user's hand when a color-coded halo surrounds it signaling to the user with which hand it can be moved with i.e. blue or red for the left and right hand respectively, see Fig.~\ref{fig:markers_creation}. If the user selects the rulers' markers to connect it with another one, the halo around the marker becomes green. Hence, each action executed as a response to the user's gestures is forthwith signaled back to the user in a visual form i.e. causes an instant change in the visualization.

\item \textbf{Photogrammetric model:} The model (C) responses instantly to all operations initiated with the user's gestural input. These include the object rotation, resizing and displacement. The first two can be seen in Fig.~\ref{fig:models_rotation}- \ref{fig:models_resizing} whereas the gesture used to grab and change position of an object is shown in Fig.~\ref{fig:models_movement} respectively. As such, all these changes in the visualization are instantly presented to the user but are not signaled to the hand-tracking sensor (B) as there is no need to use this information by this element.

\item \textbf{Measurement tools:} The tools include the rulers (vectors/lines in the 3D space), ruler's markers and the legend show to the user through the heads-up display (HUD). These can be seen in Fig.~\ref{fig:placing_markers}-\ref{fig:retaking_measurements}. All of these respond to actions invoked with the user's hands. The rules' markers can be created, removed or displaced and highlight while the user is gazing over them with a cross-hair. Moreover, their halo signalizes to the user which hand they could be moved with, i.e. blue or red for the left and right hand respectively, and green whilst the user is in the process of connecting them with a ruler. The rulers themselves change their length and position depending on the user's displacement of the attached markers and can be completely removed from the visualization as well. The legend itself responds to the changes made over the rulers showing the most recent state of the taken measurements. Hence, all these elements are sending unidirectional visual signals to the user (A) and respond to unidirectional signals received using hand-tracking (B). 
\end{enumerate}

\subsection{Visualization Framework: VR Environment}
To implement and test our system we used the Unity \cite{unity} game engine together with a small number of Unity assets \cite{unity_vr_samples_pack, unity_off_screen_arrows}. The hardware setup consisted of the Oculus Rift HMD\cite{oculus_utilities, oculus}, Oculus position-tracking sensor and the Leap Motion \cite{leapmotion} hand-tracking sensor. All these three elements can be seen in Fig.~\ref{fig:oculus_gear}. The VR-based visualization framework itself that is underpinning our tool was previously developed for the application to aeronautical design i.e. the \textbf{AeroVR} system described in detail in series of publications (see Tadeja et al.\cite{tadeja_aiaa_2019, tadeja_chi_2019, tadeja_rae_2020, tadeja_aiaa_2020}). However, the interaction method has been vastly changed and amended with the hand-tracking capabilities provided by the Leap Motion \cite{leapmotion} sensor and developed around the Leap Motion SDK \cite{leapmotion}. This subsection will describe in detail all the key elements of our framework.

\subsubsection{Photogrammetric Model: Pipe-Rack}
\begin{figure}[!ht]
    \centering
\includegraphics[width=\textwidth]{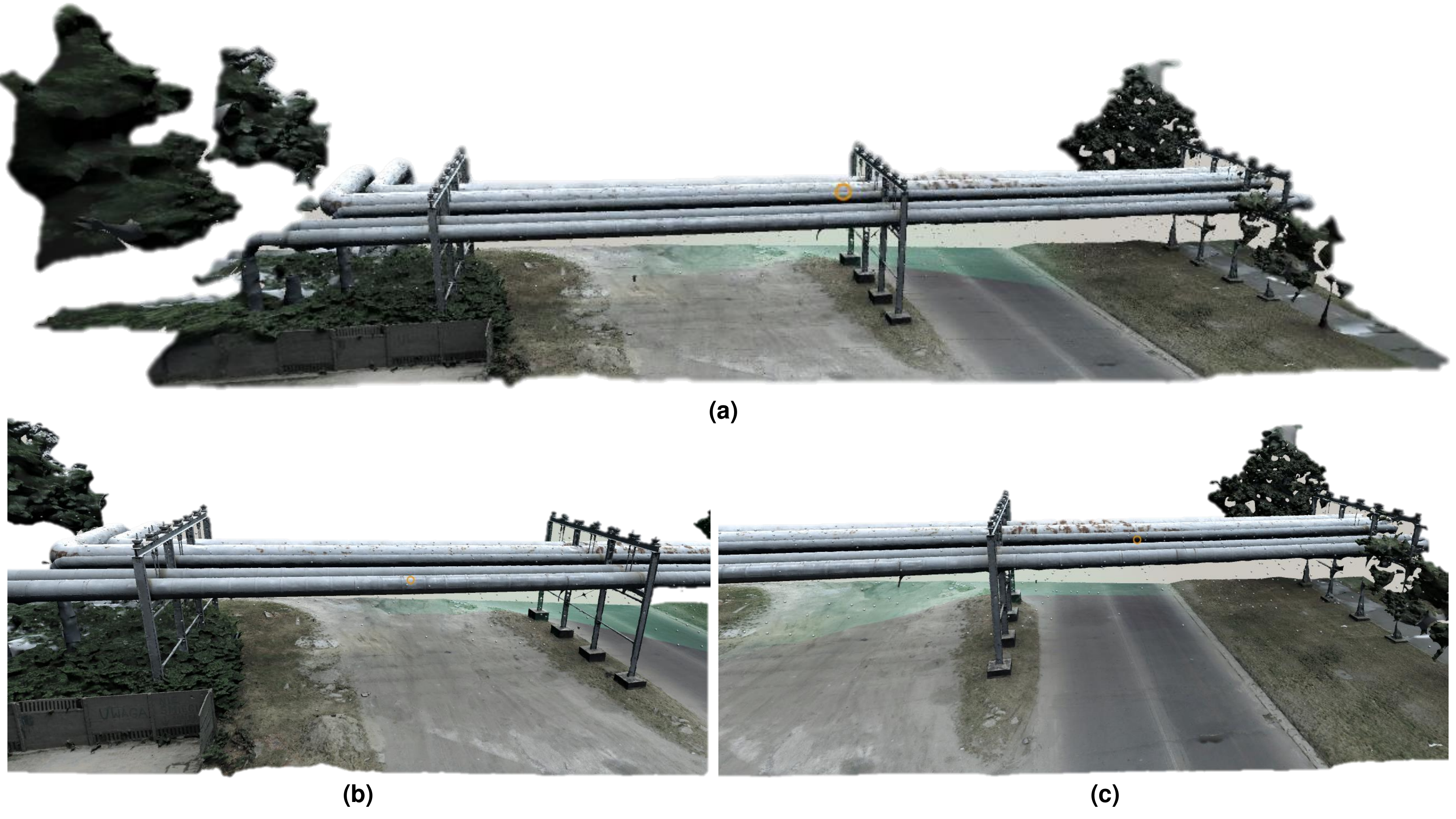}
    \caption{The 3D photogrammetric model of the pipe-rack \cite{Estakada_Model} as seen by the VR user: (a) the full model; (b) zoom-in on the left-hand side of the model; (c) zoom-in on the right-hand side of the model. The user's gaze direction is marked with the orange cross-hair.}
    \label{fig:pipe_estacada_model}
\end{figure}

The pipe-rack is located in Lodz, Poland, at the junction of the Lodowa and Andrzejewskiej streets \cite{Estakada}. The data used to reconstruct this installation, in the form of a 3D photogrammetric model seen in Fig.~\ref{fig:pipe_estacada_model}, was captured by an unmanned aerial vehicle (UAV). The model is available at \cite{Estakada_Model}. Data was processed by a commercial package - ContextCapture \cite{ContextCapture}. In this case, we can define the process as multi-image close-range digital photogrammetry (due to the source of data and the distance to the physical object, even though the data was acquired using an UAV). The camera-capture photographs were obtained during a 10 minutes UAV flight. The model was reconstructed from 204 photos with re-projection error (RMS) of 0.58 pixels and was geo-referenced using a photo meta-data gathered by the UAV and then verified in the pre-processing phase of the process of 3D model creation. As a result, the model was properly scaled and geo-positioned. The flight trajectory was a combination of an automated mission programmed with Drone Harmony software \cite{DroneHarmony}, with an additional manually steered flying passes.

Using UAVs to acquire data for photogrammetry models that are later on included in an immersive environments, as remarked by Huang \cite{huang_pieces_2019}, seems to be one of the most preferred methods, especially when dealing with larger structures or areas. Furthermore, Xu et al.~\cite{xu_skeletal_2016} reflects, that Structure-from-Motion (SfM) algorithms that can use, for example, UAV captured multi-view image data in order to reconstruct 3D scenes, have a range of application, including heritage documentation, urban planning and VR \cite{xu_skeletal_2016}. For instance, Lee et al.~\cite{lee_analysing_2018} used drones to capture data before placing the resulting models in a VR interface.

These particular pipelines are used to transport the heating medium to the residential buildings and industrial facilities located in this district of the city. The piping system section presented for visual inspection in this model is supported by three steel frames in order to safely lead pipes above the road. Supporting system need to restrict the deflection caused by gravity loads as well as wind and other environmental factors. Such pipelines and supporting frames need to be inspected on a regular basis. On the other hand, numerous difficulties and threats for inspecting person exists. With the help of a UAV-assembled camera only, this procedure might be executed faster and safer, but not necessarily precisely. However, having a correctly dimensioned 3D photogrammetric model, created with a software allowing multi-image digital photogrammetry, and enhancing the physical object reconstruction procedures with automatic algorithms for camera positioning and triangulation, gives more time and comfort for proper and, what is important, detailed inspection in the off-line mode. The model is also a valuable resource for later referencing.

\subsubsection{Interaction Methods: Gaze and Hand Tracking}
\begin{figure}[!th]
    \centering
\includegraphics[width=\textwidth]{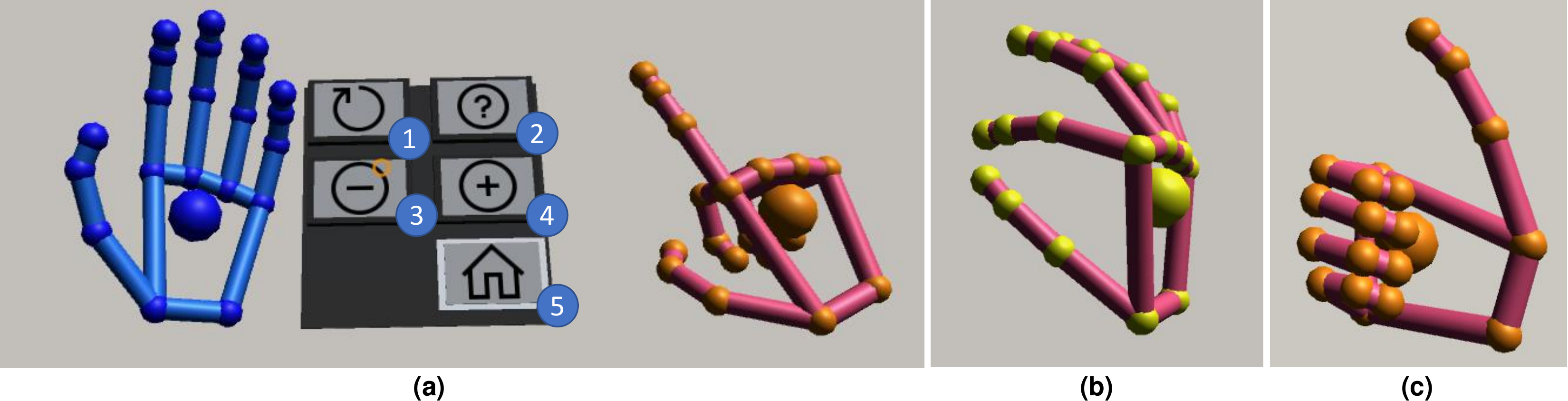}
    \caption{The avatars of the hands as seen by the VR user: blue for left-hand and red for the right-hand. (a) Making the left-hand palm-up gesture causes the attached menu to pop up. The user can press one of the buttons using any finger on the other hand. The menu options are as follows: (1) reset the visualization; (2) show the help menu; (3) switch mode to remove the rulers' markers; (4) switch mode to add more rulers' markers; (5) switch mode to manipulate the 3D model. If none of the buttons is active (pressed) the user can connect the markers with rulers to take measurements. (b-c) The \textit{pinching} and \textit{thumbs-up} gestures respectively made with the user's right hand. The (b) pinching gesture is the most basic gesture used throughout the system and can be used to, for instance, placing the model in any part of the 3D space available to the user. The icons used in the menu are courtesy of Icons8\cite{icons8}.}
    \label{fig:pinch_gesture}
\end{figure}
The manipulation of the object occurs via the user's gestural input, i.e. the hand-tracking facilitated with the Leap Motion \cite{leapmotion} sensor. The gesture-recognition capabilities were built around the Leap Motion SDK \cite{leapmotion}. Essentially, there are four types of gestures that are recognized by our system that leads to different actions depending on the mode in which the system is operating. These are (1) the \textit{left-palm up} gesture which invokes the menu as can be seen in Fig.~\ref{fig:pinch_gesture}(a) together with (2) the \textit{pointing finger} used to press one of the menu buttons; (3) item as seen in Fig.~\ref{fig:pinch_gesture}(b) shows the \textit{pinch} gesture whereas (4) the \textit{thumbs-up} gesture is presented in Fig.~\ref{fig:pinch_gesture}(c). Recognition of these gestures coupled with the use of gaze-tracking (simulated with an orange cross-hair) for selecting objects which user wants to interact with, was the basis of the main interaction method incorporated in our system. Slambekova et al.~\cite{slambekova_gaze_2012} remarks that the combination of these two techniques may have a positive impact on the tasks performed in a virtual setup.

\subsubsection{Model Manipulation Methods}
The system supports three main methods of manipulating the given 3D model be it a CAD drawing or a photogrammetric model as it was in our case: (1) displacement; (2) rotation and (3) resizing. To be able to manipulate the model, the user has to press the $[house]$ button on the left-hand menu as shown in Fig.~\ref{fig:pinch_gesture}(a). Moreover, all these actions are fluent and can be executed in unison, hence the user can simultaneously move, rotate and resize the model. Thus, making the manipulation of the model very easy and intuitive as observed during the user study.

\begin{enumerate}[(1)]
\item \textbf{Displacement}: The user can move the model into any position in the 3D space by using any hand to make a pinching gesture (i.e. grabbing the object) and moving the hand which results in the displacement of the model corresponding to the hands' movement (see Fig.~\ref{fig:models_movement}). This procedure can be repeated by the user as many times as needed to place the model in the desired location. In the Fig.~\ref{fig:models_movement} the user's avatar is included to show that the user is not changing position and the model is not scaled down but the user's perspective with regards to the model changes, hence the model seems to be smaller.

\begin{figure}[!ht]
    \centering
\includegraphics[width=\textwidth]{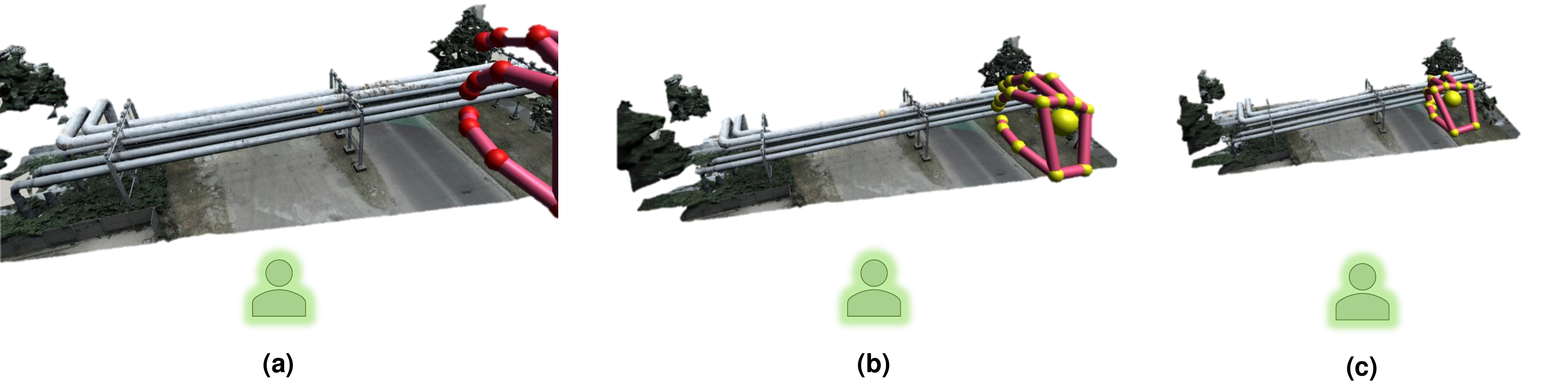}
    \caption{The user moves the model further dawn from its current position by (a) grabbing the object with the pinch gesture and (b-c) extending the grabbing hand forward. This operation pushes the model on the trajectory corresponding to the user's hand movement e.g. by straightening the user's elbow. The displaced object is not scaled but seems smaller due to change in perspective and the distance from which the user is seeing the model.}
    \label{fig:models_movement}
\end{figure}

\item \textbf{Rotation}: The rotation of a model has to be executed using both hands. When the user makes the pinch gesture with both hands and starts to rotate the hands, one of the hands can become the rotation axis while the other is responsible for the rotation arc in the $X-Z$ plane. An example of the whole process is shown in Fig.~\ref{fig:models_rotation}. Rotation in only one chosen plane at a time is also available in professional CAD software packages, as Microstation \cite{Microstation}, SolidWorks \cite{SolidWorks}, Catia \cite{Catia}. In further versions of PhotoTwinVR, adding rotation in other planes is to be added.   

\begin{figure}[!ht]
    \centering
\includegraphics[width=\textwidth]{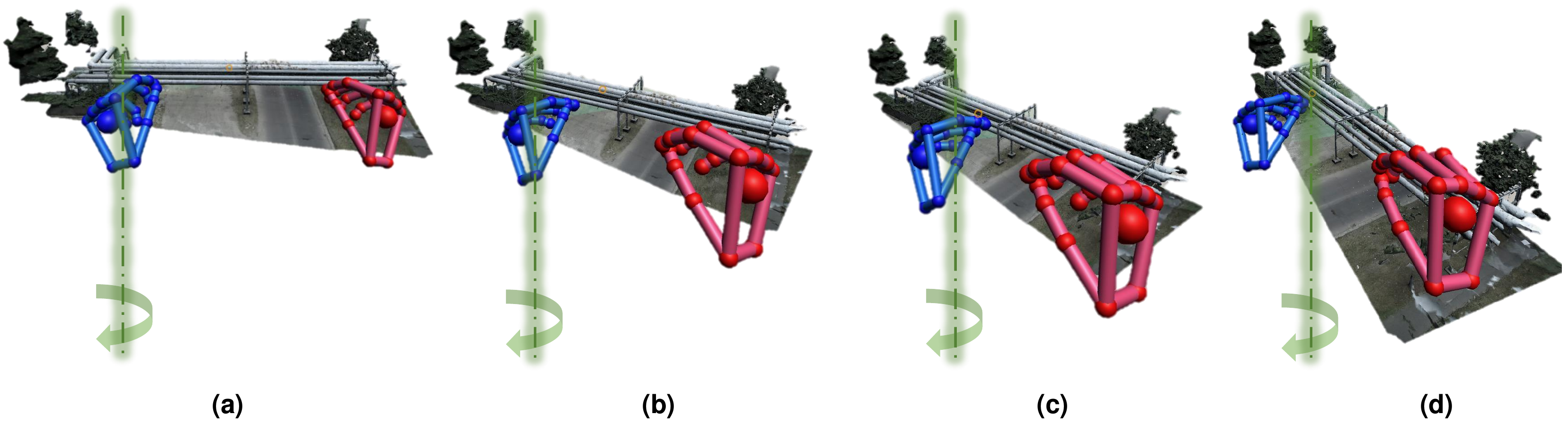}
    \caption{The user rotates the model using both hands. In this case, the left hand (blue avatar) becomes the axis about which the rotation occurs. We do not explicitly need one hand to be the axis, the virtual axis can sit anywhere on the line connecting two pinch points. When the user grabs the model (pinch gesture) with the right-hand (red avatar) and continuously moves the hand on an arc the models rotate in the $X-Z$ plane. The green arrows show the rotation direction whereas the rotation axis is shown as a dashed line.}
    \label{fig:models_rotation}
\end{figure}

\item \textbf{Resizing}: Similarly to the rotation, to resize i.e. to enlarge or decrease the model's size, the user has to grab the model using both hands. Then, if the user moves the hands apart, the model automatically scales up as can be seen in Fig.~\ref{fig:models_resizing}(a-b). In response to the opposite movement i.e. when the user moves the hands closer together, the model scales down as shown in Fig.~\ref{fig:models_resizing}(c).

\begin{figure}[!ht]
    \centering
\includegraphics[width=\textwidth]{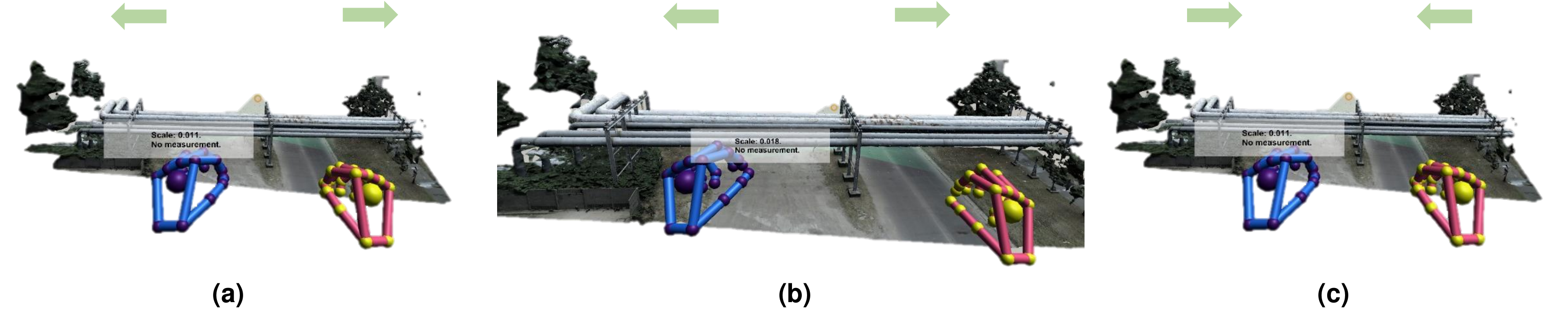}
    \caption{The user resizes the model by making simultaneous pinch gestures with both hands (blue and red hand avatars) and continuously moving the hands (a-b) apart to enlarge or moving them (c) closer together to reduce the size of the model. The green arrows show the direction in which the user's hands are moving. The heads-up display (HUD) is showing the difference in the models scale going from (a) $0.011$ to (b) $0.018$ and back to (c) $0.011$.}
    \label{fig:models_resizing}
\end{figure}
\end{enumerate}

\subsubsection{Snapping Grid}
\begin{figure}[!ht]
    \centering
\includegraphics[width=\textwidth]{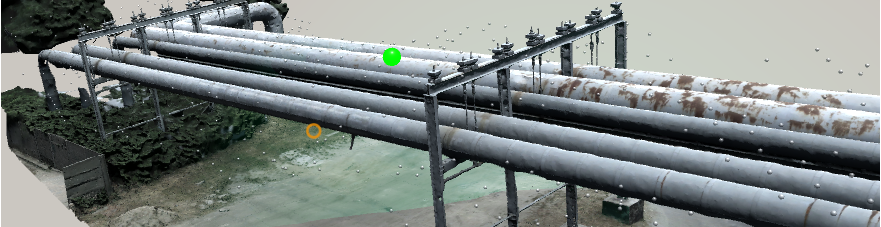}
    \caption{The snapping points (i.e. the small, grey spheres) belonging to the automatically generated snapping-grid. If the user places the ruler's marker (green spheres) in close vicinity to such a point, the marker will immediately snap into that position.}
    \label{fig:snapping_grid}
\end{figure}

The main function of the snapping grid was to support the user in the process of placing the rulers' markers around and on the model. Initially, the snapping points, i.e. small, grey spheres, are placed in preselected positions around the model. When the user tries to place the marker in close vicinity of such point, depending on the snapping radius, the marker would ``snap'' automatically to the position of the snapping point. Hence, allowing the user to place the markers quickly in the selected positions. The generation of the snapping grid was twofold: (1) we created a dense, 3D snapping grid around the photogrammetric structure, to insert a number of snapping points on it; (2) then, we checked with which such snapping-points the object collided with and removed all the points where there was no collision recorded. This process was relatively efficient as we did not have to use additional code and could rely on features natively built-in into Unity \cite{unity} game engine. We could also set the size of small, grey spheres that acted as our snapping points (see Fig.~\ref{fig:snapping_grid}) as well as the step-size used to generate the grid. The snapping radius i.e. distance from which the rulers' markers would be snapped into the position of a snapping point was automatically scaled concerning the structures current size.

\subsubsection{Taking Measurements of the Model}
\begin{figure}[!ht]
    \centering
\includegraphics[width=\textwidth]{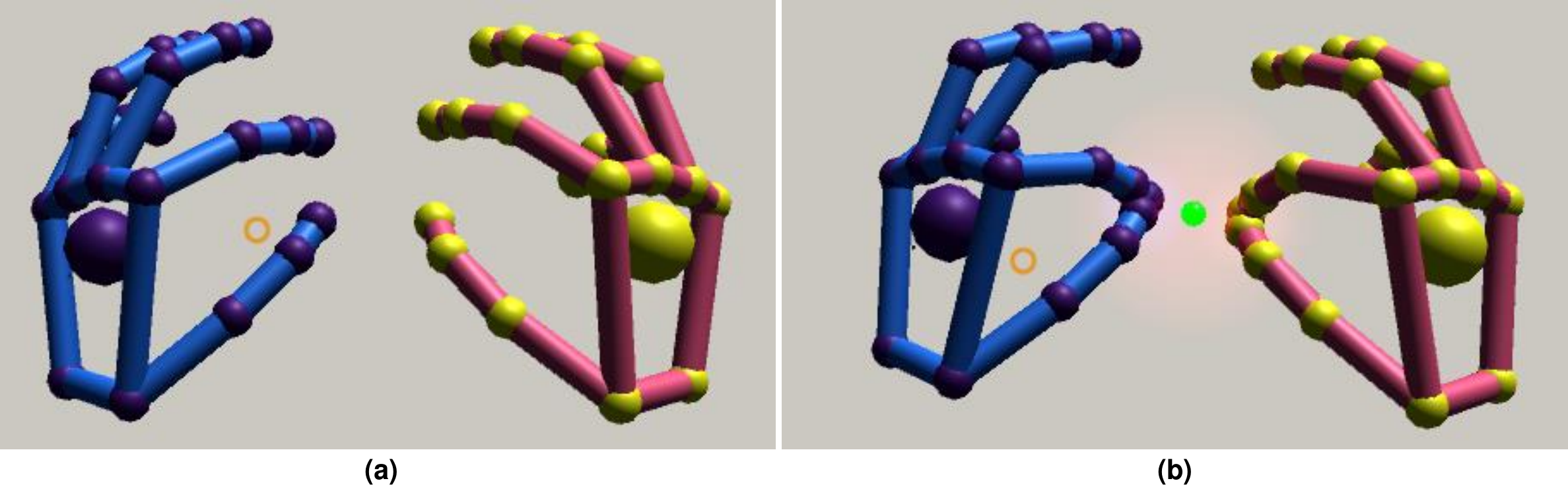}
    \caption{To generate the new rulers' markers, the user has to make a pinching gesture with both hands (a) with the all four pinching fingers at once i.e. the thumbs and pointing fingers close to each other. Then, a new marker will be generated between the user's hands (b) .}
    \label{fig:markers_creation}
\end{figure}

\begin{figure}[!ht]
    \centering
\includegraphics[width=\textwidth]{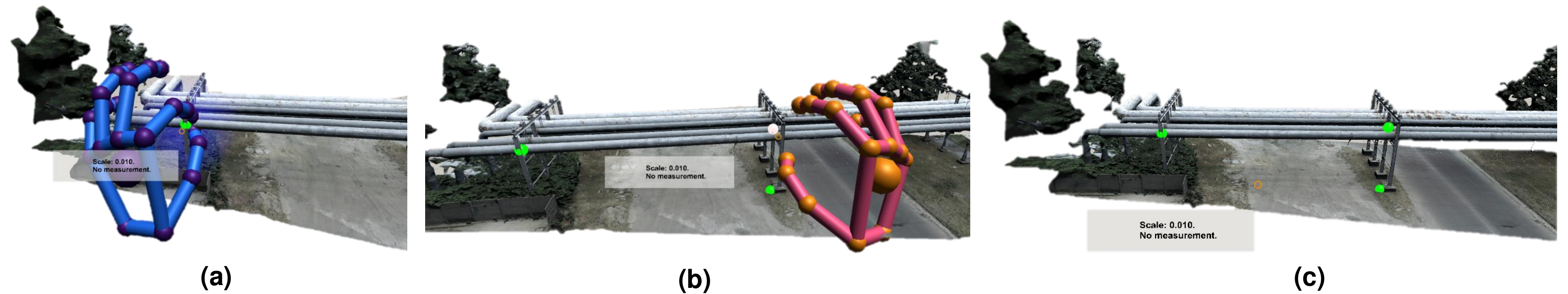}
    \caption{After creation of the ruler's markers (see Fig.~\ref{fig:markers_creation}), the user can placed them anywhere in the 3D space (c) using either hand (a-b) as long as the working mode is set to the markers generation state invoked by the push of $[+]$ button on the left-hand menu (see Fig.~\ref{fig:pinch_gesture}). If the user places the ruler's marker (green spheres) in close vicinity to such point, the marker will immediately snap into that position overlying this point (see Fig.~\ref{fig:snapping_grid}). The floating text-box HUD (Heads-Up Display) shows the current scale of the model and the information about up to three of the latest measurements whereas all the previous ones are saved automatically in a text file. To let go of a marker, i.e. release the hand's handle over it, the user has to make a thumbs-up gesture with the hand with which the marker was manipulated with or wait a two-seconds period for it to automatically deselect.}
    \label{fig:placing_markers}
\end{figure}

The process of taking a measurement, i.e. a length between any two points in the 3D space, is twofold. First, the user has to switch to the generation mode by pressing the $[+]$ button on the left-hand menu (see Fig.~\ref{fig:pinch_gesture}(a)) and generate (see Fig.~\ref{fig:markers_creation}) and place at least two rulers' markers, represented by green spheres, in the desired locations in the 3D space (see Fig.~\ref{fig:placing_markers}). Second, the user has to connect the selected markers with the ruler i.e. a vector in the 3D space. 
The length of this ruler measured from within the centers of the markers is the euclidean distance calculated in the 3D space.
Each marker can be connected to three other markers, hence, the user can, at least in principle, cover and measure with rulers the lengths of all the edges of a triangular mesh. 

\begin{figure}[!ht]
    \centering
\includegraphics[width=\textwidth]{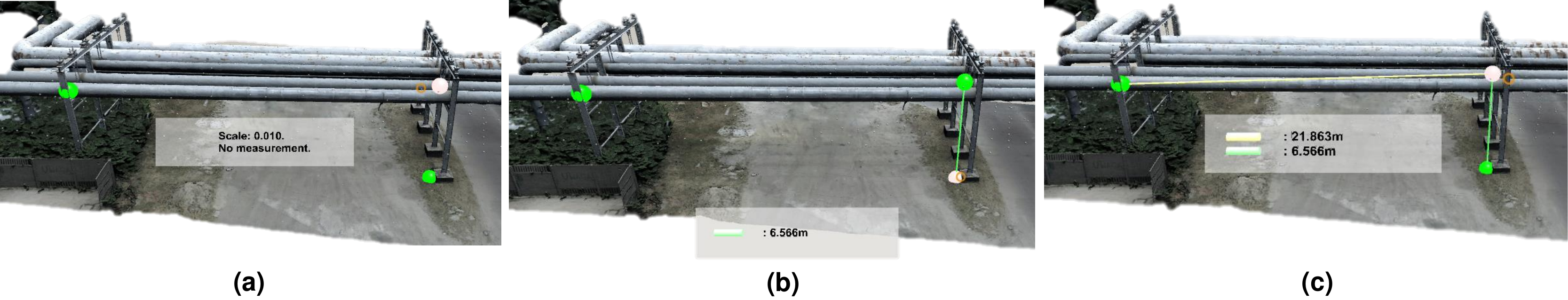}
    \caption{To measure the object's or structure's dimension one has to place two ruler's markers in the 3D space and connect them with a ruler. The length of such a ruler is equal to that dimension measured. (a) The user creates a connection between two points by gazing over (orange cross-hair) the initial marker which automatically highlights itself (top-right marker) and making a pinch gesture (see Fig.~\ref{fig:pinch_gesture}). Before any rules are added into the scene, the HUD shows only the current scale of the model (b) When user gazes over another marker and repeats the pinch gesture a new ruler is generated between the two markers (b-c). Each taken measurement is automatically written to an output file that keeps track of user actions in order of execution.}
    \label{fig:taking_measurements}
\end{figure}

\begin{figure}[!ht]
    \centering
\includegraphics[width=\textwidth]{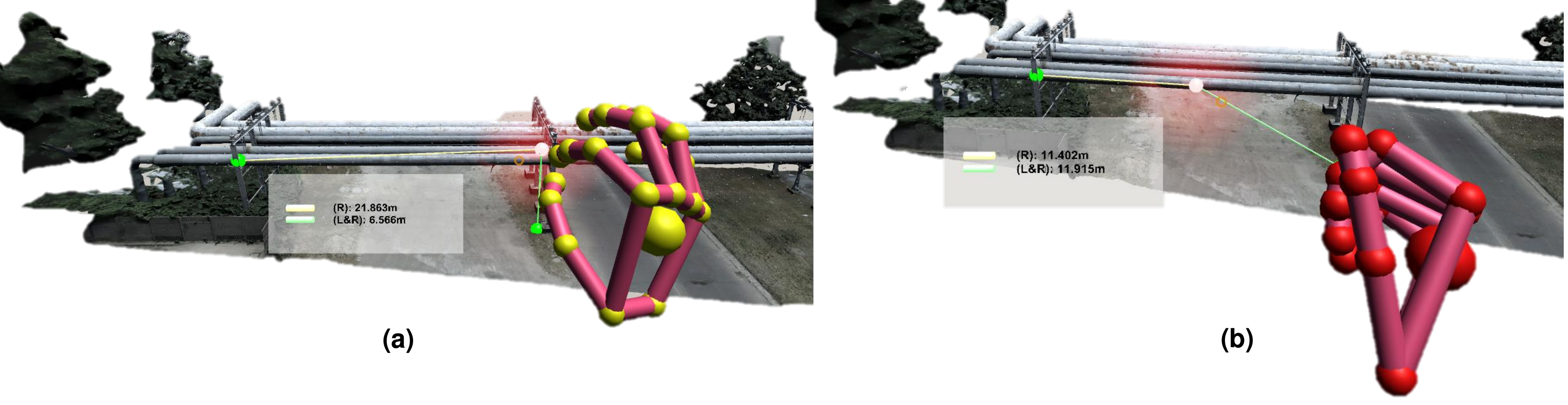}
    \caption{Once the markers are connected with rulers, when the user changes the position of any of the connected markers (a-b) the rulers will not disconnect itself from the markers and will adjust their position and length concerning the new marker's position. They will automatically recalculate their respective lengths which will be immediately visible on the user's HUD.}
    \label{fig:retaking_measurements}
\end{figure}

To not occlude the user's field of view, only the last three changes i.e. new measurements or change in the previously taken ones are shown to the user via a heads-up display (HUD) (see Fig.~\ref{fig:taking_measurements}-\ref{fig:retaking_measurements}). However, all measurements are automatically written into a file in order of when they were taken or changed.

\section{Interface Verification}
Our approach to verify if the interface proposed by us is meeting requirements and usability standards were twofold. Firstly, we used the well known Nielsen's heuristics \cite{nielsen_usability_1994, nielsen_enhancing_1994} to asses and reason about the usability of our system. Secondly, we ran a small formative study with three volunteers from the UDT \cite{urzaddozorutech} organization to test our system with domain-experts who are potential target users of such toolkit. Results of this evaluation can be found in the next two subsections. 

\subsection{Interface Usability}
\begin{enumerate}[(1)]
\item \textbf{Visibility of system status}: User hand-gestures are shown to the user in the form of hand-avatars (see Fig.~\ref{fig:pinch_gesture}). If the avatars are not within the user's field of view it means that they are outside of the sensor's field of tracking. The system continuously responds to all user actions. For instance, when the user is manipulating the model by displacing, rotating or re-sizing it, all the changes made to the model are immediately visible to the user. With respect to taking measurements of the model, all the user's actions such as generation of the rulers' markers (see Fig.~\ref{fig:markers_creation}), their placement or change thereof (see Fig.~\ref{fig:placing_markers} and Fig.~\ref{fig:retaking_measurements}) together with connecting them with the rulers (see Fig.~\ref{fig:taking_measurements}) are also immediately visible to the user. The recently taken measurements itself i.e. the Euclidean lengths of the rulers are shown to the user via the self-updating HUD (see Fig.~\ref{fig:taking_measurements}).

\item \textbf{Match between the system and the real-world}: The system supports the usage of the 3D photogrammetric model of a real-life, existing piping system (see Fig.~\ref{fig:pipe_estacada_model}) that mimic, in great detail, the original structure. Hence, the match between this main element of the visualization and the real-world is extremely high. Moreover, the user's hands were also visible to the user via the hands' avatars (see Fig.~\ref{fig:pinch_gesture}) and all used gestures such as \textit{pinching}, \textit{thumbs-up} or \textit{pointing finger} are well known and very natural to the users as well. Furthermore, technique used for taking the models measurements including manipulation of the rulers' markers i.e. using the user's hand to grab and place them (see Fig.~\ref{fig:taking_measurements}), or showing the measurements' results in the form of a map legend displayed on a HUD (Fig.~\ref{fig:retaking_measurements}) are also closely following real-world conventions. 

\item \textbf{User control and freedom}: It is very easy for the user to manipulate the model i.e. the models size, position or rotation (see Fig.~\ref{fig:models_movement}-\ref{fig:models_resizing}). Moreover, the user can change the position of the existing rulers' markers (see Fig.~\ref{fig:retaking_measurements}). The user can also completely remove any marker by pressing the $[-]$ button on the left-hand menu and gaze over the marker and simultaneously making the double-pinch gesture for the marker to be removed. This operation will automatically remove also any existing connection between the deleted and any other markers. Also, whenever necessary, the user can choose to reset the visualization to its default state by pressing the $[reset]$ button on the left-hand menu as shown in Fig.~\ref{fig:pinch_gesture}(a). This action will preserve all the previously taken measurements in the form of a text file.

\item \textbf{Consistency and standards}: In some respect, our interaction methods follows standards known from other systems. For instance, the user has to push the hands apart or bring them closer together, respectively, to increase or decrease the model's size. The same can be said about placing the rulers' markers (see Fig.~\ref{fig:placing_markers}) which follows grabbing and dropping gestures that the user would have to execute to place a marker in the real world as well.

\item \textbf{Error prevention}: The situation in which the user could encounter an error was to place the ruler's markers in a way that would cause vastly imprecise measurements of an object. We tried to remove completely such situations by automatically placing the snapping grid as shown in Fig.~\ref{fig:snapping_grid}. However, the used approach, at least in the current version of the code, proved not to be precise enough for most users. In the future iterations, we will design a more robust algorithm that would allow the system to automatically placing the grid's snapping points in most desirable locations around the model, for instance on the vertices given by its triangular mesh. 

\item \textbf{Recognition rather than recall}: First, as we are using three-dimensional photogrammetric models of real-life objects, (see Fig.~\ref{fig:pipe_estacada_model}) they are, in a way, intuitively recognizable by the domain-expert users as objects with which they are working daily within a physical environment. Moreover, the users can immediately recognize hands-avatars and easily perform the set of four gestures (see Fig.~\ref{fig:pinch_gesture}) identifiable by the hand-tracking sensor \cite{leapmotion}. Presenting the final measurements in a form a legend similar to the ones encountered on the cartographic maps.

\item \textbf{Flexibility and efficiency of use}: The majority of the user's actions can be carried out by both left and right-hand user alike without any changes made to the system. The hand-tracking sensor \cite{leapmotion} recognize the gestures (see Fig.~\ref{fig:pinch_gesture}) made with any or both the hands.

\item \textbf{Aesthetic and minimalist design}: The visualization consists only of three elements: (i) the main visualized object, in this case, the pipe-rack model; (ii) the user's hands-avatars; (iii) the rulers' markers, rulers and their legend displayed on a HUD. The system does not visualize any other additional information with except the left-hand menu that can be called and hidden on the user's demand with a simple hand gesture as can be seen in Fig.~\ref{fig:pinch_gesture}(a).

\item \textbf{Help users recognize, diagnose, and recover from errors}: The main error that user can make is to place the ruler's markers in an unwanted position or connect wrong markers which will be immediately visible to the user on inspection of the model. All these operations can be undone and, if needed, the user can reset the entire visualization to its default setting simultaneously preserving obtained measurements in an output file.

\item \textbf{Help and documentation}: The system has built-in help that can be accessed by simply pressing the $[?]$ (help) button on the left-hand menu as shown in Fig.~\ref{fig:pinch_gesture}(a).
\end{enumerate}

\subsection{Hand-Menu Usability}
With regards to the hand-attached menu, Azai et al.~\cite{azai_open_2018} investigated placement of an immersive application menu in front of the user’s palm, and based on the study and participant's comments suggested that such menus are feasible when tasks require frequent change in menu positioning and for which the menu does not have to be constantly displayed \cite{azai_open_2018}. 
Such a situation closely relates to our system as well. Moreover, the authors proposed a set of design guidelines that, for instance, suggest that vertical layout is more favorable and three columns and four rows of items is a feasible limit for such menus \cite{azai_open_2018}. In our case, the menu had two columns and three rows in total as shown in Fig.~\ref{fig:pinch_gesture}(a). 

With respect to the pinch gesture, user study conducted by Jude et al.~\cite{jude_grasp_2016} concluded that the users favor the \textit{grasp} over \textit{pinch} gesture which in turn was preferred over the \textit{grab} gesture. However, as our system very poorly differentiated between the \textit{grasp} and \textit{pinch} gestures as, for instance, seen in Fig.~\ref{fig:models_rotation}, thus, we are allowing the users greater freedom in the selection of their favorite manipulation technique. 
Cui et al.~\cite{cui_understanding_2016} investigated mental models accompanying mid-air gesture-based interaction facilitated with Leap Motion \cite{leapmotion}. The authors note that the users preferred bimanual interaction and having virtual hands assist the users in assessing respective virtual object's sizes \cite{cui_understanding_2016}. Furthermore, it helps the user in planning and execution of a composite, complicated task \cite{cui_understanding_2016}.

\subsection{Formative Study}
To conduct our experiment, we chose to apply the observational study methodology cataloged by Lam et al.~\cite{lam_empirical_2012} as one of the methods often used for visualization evaluation. Two models were used to prepare and perform the formative study, including the training video presented to the volunteers before they participated in the experiment. After seeing the short video clip, the participants were asked to perform slightly different tasks depending on with which model they were working with. Here, we report only about the results related to the pipe-rack model. This model, shown as it is seen by the VR headset wearer can be seen in Fig.~ \ref{fig:pipe_estacada_model}. 

\subsubsection{Study Participants}
All the participants, hereinafter referred to P1, P2 and P3, have years of experience working for either the industry (P3) or the \textit{Urzad Dozoru Technicznego} \cite{urzaddozorutech} (UDT, \textit{eng.} Office of Technical Inspection) \cite{urzaddozorutech} which is an EU Notified Body No. 1433 (P2-P3), purpose and work of which is described in detail in Section 1.

The participants were asked to fill in the \textit{Simulation Sickness Questionnaire} (SSQ) developed by Kennedy et al.\cite{kennedy_simulator_1993} and restructured by Bouchard et al. \cite{bouchard_revising_2007} before and after they spent time in the VR environment during the experimental phase of the study. They were also told to immediately discontinue the experiment if they developed any of the simulation sickness symptoms or other forms of discomfort. Moreover, we asked them to fill in the \textit{Flow Short Scale} (FSS) \cite{flowshortscale}, \textit{Igroup Presence Questionnaire} (IPQ) \cite{igroup_presence} and \textit{NASA Task Load Index} (NASA TLX) \cite{hart_development_1988}. The first two were used to gain input of how participants experienced the VR environments and the latter to check their cognition load whilst executing the given task. All the above questionnaires were translated into the Polish language by the authors. Also, we obtained an independent validation of our translations from a native-speaking researcher from an uninvolved academic institution with expertise in both psychology and human-computer interaction research. Before commencing the experiment, using a Likert-like scale, the participants were also asked for complete self-assessment of their level of familiarity with the VR environment and hand-tracking sensor such as the LeapMotion \cite{leapmotion}.

\textbf{Participant 1} (\textbf{P1}) was 26 years old and holds an engineering degree in power engineering. The professional experience of this Participant included three years of working for the UDT, mainly as an installation inspector responsible also for the analysis of gathered measured data both during the inspection and installation operations.

\textbf{Participant 2} (\textbf{P2}) reported being 41 years old and holds a master and an engineering degree in process engineering. He worked for six years as a plant pipelines designer. He also reported to have vast experience with a range of professional designing and CAD modeling software.
 
\textbf{Participant 3} (\textbf{P3}) was 29 years old working and holds an engineering degree in power engineering. He has four years of experience as a plant pipelines designer and has worked with many software packages considered as standards in the industry. 

\subsubsection{Training and Experimental Procedure}
As mentioned before, our participants reported very basic or non-existent prior experience with using either the VR or hand-tracking interfaces. Hence, some initial training was required to get them acquainted with our system and its capabilities.

Before the experimental part of the study, each of the three volunteers was given short oral instructions and presentation of our tool. The showcasing had the form of a short video clip with subtitles in the participants' native language, lasting 3:18 minutes and presenting details of all the possible interaction techniques. The video was featuring two photogrammetric models including the pipe-rack as one of them.  

In the experimental part the participants were asked to carry all the work while standing and instructed to first play with the system for a while to familiarize themselves with the model's manipulation method, and then to fulfill the following task (translated to the English language): \textit{Please measure -- using the interaction methods presented -- the height of the supporting structure} [part of the pipe-rack model (see Fig.~\ref{fig:pipe_estacada_model})] \textit{and the distance between neighboring structures, as well as the width of one of them. Please take all other measurements that you consider relevant concerning surveying the asset.} They were also given unlimited time to perform the given task. Moreover, they were asked to instantly share aloud their thoughts during the experiment, and inform the researcher about what and why they are trying to achieve at the moment. Also, they were asked to inform the researcher what obstacles they are facing, as well as comment on what they find frustrating and, in positive meaning, what works well for them and what makes them satisfied from the tool they were using. The researchers were tracking the participants' actions and their respective results in a real-time on a computer screen where the participant's field of view was continuously streamed. The experiment was followed by a semi-structured discussion between the participant and the researchers. The experimental phase was video and audio recorded for further analysis.

These exercises aimed to give the participants a possibility to learn the interaction methods while simultaneously giving them a goal on which they could focus. This was especially important as we wanted to observe their behavior when using our system with context to their engineering background, training and experience. We wanted to see how they are going to approach learning the interaction method and how easy or how difficult they would find the entire process. In regard to this, we were not interested in very detailed measurements but rather about the overall participants' experience and observed behaviour. As such, both parts of the given exercise were considered as finished when either the researcher or participant decided that they had gained enough fluency in using the system so the participant would be able to carry out any measurements of the given model. Bearing this in mind, the participants were using the headset for approximately: P1-31.35 min, P2-21.32 min, and P3-14.67 min. 

\subsubsection{Questionnaire and Survey Results}
\begin{table}[]
\centering
\begin{tabular}{@{}|c|c|c|c|c|c|c|c|@{}}
\toprule
\multicolumn{2}{|c|}{}                                                 & \multicolumn{6}{c|}{\textbf{Questionnaire and Survey Results}}                                                                 \\ \midrule
\textbf{Participant} & \multicolumn{1}{l|}{\textbf{Exercise Duration}} & \textbf{IPQ}           & \multicolumn{2}{c|}{\textbf{SSQ}}      & \multicolumn{2}{c|}{\textbf{FSS}} & \textbf{NASA TLX}    \\ \midrule
\textit{number}      & \textit{minutes}                                & \textit{average score} & \textit{nausea} & \textit{oculo-motor} & \textit{flow}  & \textit{anxiety} & \textit{total score} \\ \midrule
P1                   & 31:35                                           & $3.21/7$               & $0/27$          & $0/21$               & $3.4/7$        & $4.0/7$          & $10/100$             \\ \midrule
P2                   & 21:32                                           & $3.64/7$               & $3/27$          & $3/21$               & $5.4/7$        & $3.0/7$          & $28/100$             \\ \midrule
P3                   & 14:67                                           & $3.14/7$                 & $1/27$          & $3/21$               & $5.5/7$        & $4.5/7$          & $20/100$             \\ \bottomrule
\end{tabular}
 \caption{The table shows the results of the questionnaire and survey that each participant had to undergo prior (only SSQ) and after carrying out a task in the VR environment. The time duration is given only as an approximation as the exercise given to the participant had no strict time constraints.}~\label{tab:quest_results}
\end{table}
Since the number of participants was fairly small i.e. the gathered sample size is not large enough to warrant any statistical analysis, here we present the results of the questionnaire and survey results that can also be found in Tab.~\ref{tab:quest_results}. 

With respect to the SSQ \cite{kennedy_simulator_1993}, P1 reported none symptoms whereas P2 (3/21 and 3/27 for nausea and oculomotor strain respectively) and P3 (3/21 and 3/27 for nausea and oculomotor strain respectively) reported experiencing some slight simulation sickness symptoms. It is worthy of noting that both P1 (2/21 and 0/27 for nausea and oculomotor respectively) and P3 (2/21 and 2/27 for nausea and oculomotor strain respectively) noted about having some preexisting symptoms before the experiment commenced.

In case of the FSS \cite{flowshortscale}, all three participants reported relatively high levels of flow and anxiety: P1 (4.3/7.0 and 4.7/7.0 respectively), P2 (5.4/7.0 and 3.0/7.0 respectively), and P3 (5.5/7.0 and 4.5/7.0 respectively).

Regarding the IPQ \cite{igroup_presence}, similarly to Schwind et al. \cite{schwind_using_2019} we have averaged the results from all the questions that participants answered. With the seven-point scale in each, the highest possible score that could have been given was seven. The scores were:  P1 (3.21/7.0), P2 (3.64/7.0), and P3 (3.14/7.0) respectively. These results were to be expected as we were not designing our environment to increase the user's feeling of presence but we rather optimize for the user's comfort and ease concerning the task fulfillment.

For the NASA TLX \cite{hart_development_1988}, the participants reported relatively small levels of experienced cognition load i.e. P1-10/100, P2-28/100, and P3-20/100 respectively.

\subsubsection{Measurements Results}
The results of the measurements taken to fulfill the requirements of the first part of the exercise i.e. to measure the width and height of the supporting structure, as well as the distance between the neighboring structures are shown in Table~\ref{measurements_results}. All of the participants managed to obtain similar results with certain variability and precision within the $\pm 0.5 [m]$ meter. It should be noted that the most experienced participant, P2, stated that although it is hard to make general comments on the accuracy required in such measurements (as it depends on the engineering purpose in hand), \textit{in the case of pipelines (...) even half a meter is enough} [of accuracy] and \textit{(...) no more is needed} [based on his vast experience as an inspector, in this particular case] \textit{because it is quite a large pipe (...)}.

\begin{table}[!h]
\centering
\begin{tabular}{@{}|c|c|c|c|@{}}
\toprule
 & \multicolumn{3}{c|}{\textbf{Measurements Results: Pipe-Rack Supporting Structure}} \\ \midrule
\textbf{Participant} & \textbf{Height} & \textbf{Width} & \textbf{Distance b/w Neighbouring} \\ \midrule
\textit{number} & \textit{{[}m{]}} & \textit{{[}m{]}} & \textit{{[}m{]}} \\ \midrule
\textbf{P1} & $7.128$ & $10.443$ & $23.730$ \\ \midrule
\textbf{P2} & $7.028$ & $10.441$ & $24.044$ \\ \midrule
\textbf{P3} & $6.895$ & $10.047$ & $23.433$ \\ \bottomrule
\end{tabular}
\caption{The measurements taken by the participants. The second column from the left shows the height of the supporting structure as measured by the participant using our measurement toolkit. The third presents the measured width of the supporting structure whereas the last one shows the distance measured between the two neighbouring supporting structures. }
\label{measurements_results}
\end{table}


\begin{table}[]
\centering
\begin{tabular}{@{}|c|l|c|c|l|c|c|l|@{}}
\toprule
\multicolumn{2}{|c|}{}                     & \multicolumn{6}{c|}{\textbf{Quantitative Feedback: Model Manipulation}}                                                                                 \\ \midrule
\multicolumn{2}{|c|}{\textbf{Participant}} & \textbf{Total Displacement}    & \multicolumn{2}{c|}{\textbf{Maximum Rotation}}    & \multicolumn{3}{c|}{\textbf{Scale {[}nominal = 0.05{]}}}             \\ \midrule
\multicolumn{2}{|c|}{\textit{number}}      & \textit{nominal units}         & \multicolumn{2}{c|}{\textit{angle {[}degrees{]}}} & \textit{minimum}             & \multicolumn{2}{c|}{\textit{maximum}} \\ \midrule
\multicolumn{2}{|c|}{P1}                   & $106.16$                       & \multicolumn{2}{c|}{$153.15^{\circ}$}             & $0.004$                      & \multicolumn{2}{c|}{$1.479$}          \\ \midrule
\multicolumn{2}{|c|}{P2}                   & $12.82$                        & \multicolumn{2}{c|}{$83.48^{\circ}$}              & $0.002$                      & \multicolumn{2}{c|}{$0.050$}          \\ \midrule
\multicolumn{2}{|c|}{P3}                   & $26.40$                        & \multicolumn{2}{c|}{$78.55^{\circ}$}              & $0.004$                      & \multicolumn{2}{c|}{$0.300$}          \\ \bottomrule
\end{tabular}
 \caption{The table contains information about extents to which the participants manipulate the pipe-rack model. The second column from the left show total displacement of the model in nominal units; the third maximal angle of rotation of the model; and the fourth and fifth showing minimal and maximal scale of the model applied by the participant. For instance, at some point, P1 enlarged the model by almost 30 times. }~\label{tab:manipulation_results}
\end{table}

\section{Discussion}
When analyzing the users' comments and behavior, one has to remember that the domain-expert participants can be, in a way, largely biased by their current work interests and their professional background. Hence, for instance, their suggestions regarding what they would like to have as an addition to the presented software can vary. Here, we discuss three different areas of particular interest, i.e. (1) how they commented on and interacted with the model while immersed in our VR environment; (2) how helpful was our interface in carrying out the assigned task using the given measurement toolkit, and (3) what are their suggestions and ideas of further development directions for our system.

\subsection{Model Manipulation}
On the other hand, with respect to the model manipulation, the participants seemed to be happy with the way they could interact with the model, for instance, P1 said that \textit{(...) what I like very much is (...) rotating, moving, bringing closer (...)} [the model] and further, \textit{here I can see the entire pipeline, all suspensions but it does not give me the opportunity to focus on details.} Moreover, as can be seen in Table~\ref{tab:manipulation_results}, P1 was the one who moved the object the most and rotated it by the largest angle as well. He was also the participant who magnified the object by the largest factor, i.e. almost 30 times (see Table~\ref{tab:manipulation_results}). Concerning this, it is also worthy of noting that he was the youngest and the least work-experienced participant, which may have had influenced his behaviour. P2 commented that \textit{in terms of displacement, zooming in and out of the model it is OK}. The observation of the participants' behaviour would also suggest and confirm that all the participants quickly gain fluency in model manipulation.

\subsection{Measurement Toolkit}
One of the key parts of the system was the snapping grid, which, at least in a current form, was decisively superfluous and, instead of being helpful, it caused rather more confusion for the users. For instance, P1 commented that \textit{these spheres} [snapping points] \textit{don't help me (...) they are confusing, because they suggest something to me.} and also remarked that \textit{if it was not for these} [snapping] \textit{spheres, I could indicate more precisely the place I want} [to put the marker at]. He also suggested that it may be helpful for the user to be able to turn the snapping grid on and off. Similarly, P3 was experiencing issues with grabbing and moving the ruler's markers. However, he remarked that \textit{(...) I think it's a question of practice}.

In general, even though all the participants (P1-P3) managed to capture the required measurements, with a certain degree of variability between them, they would prefer to have a higher accuracy when placing the markers. This possibility was constrained by the snapping radius predefined in the interface program, which grew in direct proportion to the model increase in size. There are a few ideas that can be implemented in this case. For instance, the snapping radius could be constant and relatively small, or it could decrease following the model's enlargement. Moreover, as suggested by P1, it could also be turned on and off on the user's demand. The algorithm for automatic generation and placement of the snapping points could also be refined to facilitate better the user needs. For example, with the help of computer vision algorithms, the system could recognize both major and minor elements of interest, such as the supporting structure's parts and pipeline (P2), together with cracks and discolouring on the pipe (P1), and place the snapping points relatively close to these. Consequently, they could be used as indications of the crucial parts of the model, where users focus should be put. 

\subsection{Ideas For System Extension}
The discussion with P1 revealed that he would be more interested in the details (surfaces, cracks, corrosion) of the pipeline trestle bridge rather than in taking the measurements of it. He commented that P1: \textit{(...) If the details could be seen, it would be the perfect tool for me. I do not see these details, and this is a problem for me.}. For instance, he can see that there is discolouration or a crack visible on the pipe model that he would like to inspect with intense detail. He would also be interested in additional information overlaid on the model, like the temperature or pressure values at certain points of the pipe. This type of data he would like to inspect due to his current work interests. P2 commented that \textit{I am interested in the distances between supports,} [distance] \textit{between the change of direction of the pipeline, i.e. between the knees. The condensation, things like that. To draw a pipeline route.} He also remarked that system like ours might have other benefits. For instance, the engineer does not have to be physically present at the object site to take measurements, and as such, does not have to worry about the atmospheric conditions at the day of measurement.
Moreover, frequently, direct exposure to the substances in the pipes, for instance, while working at the oil refinery, can be hazardous to human health, and as such, any contact with them should be minimized or, ideally, completely avoided. In the case of our system, only a UAV or other unmanned vehicle operator would have to be physically present relatively near the object, the photographic data of which is acquired. Furthermore, once the model is ready, one can introduce the avatars of sensors placed in the same spots as on a real-life object. These could provide additional, real-life sensor data streamed directly onto the model.

All these ideas can be added on top of the existing PhotoTwinVR implementation. However, the applicability depends on the accuracy of the model itself, i.e. if it contains views of all key elements, photos are detailed enough, and the final model is properly positioned in and calibrated to the global reference system (latitude, longitude and elevation).

\section{Summary and Conclusions}
In this paper, we investigated the application of VR to unlock the full potential offered by digital 3D photogrammetric models of reality reconstruction generated using the state-of-art photogrammetric technologies. 

We identified that, to the best of authors' knowledge, there is little research published on the feasible interaction methods in the VR-based systems augmented with the 3D photogrammetric models. Especially in the context of the systems designed to help the professional to derive important information on the physical objects using gestural interfaces.

Consequently, we presented the PhotoTwinVR -- an immersive, gesture-controlled system for manipulation, inspection and dimension measurements of a 3D photogrammetric model of a physical object in the VR environment populated with a pipe-rack model \cite{Estakada_Model} obtained with the help of a commercially available software package.

To verify and validate our analysis, completeness and implementation of the system requirements, we run an observational study \cite{lam_empirical_2012} with a small group of three domain-expert participants. Notably, two of the participants were representatives of \textit{Urzad Dozoru Technicznego} (UDT, \textit{eng.} Office of Technical Inspection) \cite{urzaddozorutech} -- an EU Notified Body No. 1433. 

In the study, we explored a possibility of using the hand-tracking and gesture recognition capabilities afforded by the Leap Motion sensor \cite{leapmotion}, coupled with the gaze-tracking and the VR environment offered by the Oculus Rift VR headset \cite{oculus}. The qualitative data and our own observations of the participants behaviour and comments gathered during this experiment allowed us to reason about the feasibility of using such interaction technique to interact and extract relevant data from 3D photogrammetric models whilst being immersed in VR. The resulting measurements (see Table \ref{tab:manipulation_results}), taken during the off-line inspection of a pipe-rack (see Fig.~\ref{fig:pipe_estacada_model}) by domain-expert participants suggest that the experts were able to obtain pipe-rack \cite{Estakada, Estakada_Model} dimensions within a reasonable accuracy 

Moreover, all the participants were able to promptly gain fluency in manipulating the model with their hand gestures, as was confirmed by their comments.

With regards to future work, as suggested by Azai et al.~\cite{azai_open_2018}, we will allow the user to choose their dominant hand and as such, allow the palm-up menu to be attached to either of the user's hands based on preference. 

Furthermore, Caputo et al.~\cite{caputo_evaluation_2015} investigated different methods of 3D object's translation and rotations in VR using the Leap Motion \cite{leapmotion} controller. One of the rotation method described by the authors is the \textit{one-handed translation and second-hand rotation (2HR)}~\cite{caputo_evaluation_2015} which was later compared to results of a user study that suggested that in-direct, more abstract methods of rotation can be more preferable and beneficial to the users. Following this result, we will implement different, in-direct methods of rotation for subsequent research.

We are also planning to test these changes and updates together with ideas of further extensions proposed by the study participants in a series of controlled user experiments.

To sum up, the study revealed the PhotoTwinVR's potential to be applied in practical real-words cases, including off-line inspections of pipelines, which are common in chemical, petrochemical, power and oil\&gas industries. It also showed that there is a great promise in populating VR immersive environment with digital 3D photogrammetric models.

\section*{Acknowledgements}
The authors would also extend the gratitude to the \textit{Bentley Systems Incorporated} for allowing us to access the \textit{ContextCapture} software used to generate the 3D reality meshes of the structure used to carry out this study. The authors would also like to thank the following people: Krzysztof Kutt for verifying translations of the questionnaires, to Wioleta Caba\l a, for advice on the SSQ translation, and to Jerzy Soko\l owski and Roman G\'{o}recki from UDT \L \'{o}d\'{z} for their aid in the preparations and execution of this study. Finally, the authors would like to thank the volunteers for their participation and valuable suggestions and comments.

Funding: This work was supported by the Cambridge European \& Trinity Hall, the Tsinghua Academic Fund for Undergraduate Overseas Studies, and the Fund from Tsien Excellence in Engineering Program.

\bibliography{references}
\bibliographystyle{asmems4}
\end{document}